\documentclass[%
reprint,
showpacs,preprintnumbers,
 amsmath,amssymb,
 aps,
prb,
]{revtex4-1}

\usepackage{graphicx}
\usepackage{dcolumn}
\usepackage{bm}
\usepackage{color}

\begin{document}


\title{Commensurability oscillations in a quasi-two-dimensional
  electron gas\\ subjected to strong in-plane magnetic field}

\author{L.\ Smr\v{c}ka}\email{smrcka@fzu.cz}
\affiliation{Institute of Physics,
Academy of Science of the Czech Republic,~v.v.i.,\\
   Cukrovarnick\'{a} 10, 162 53 Prague 6, Czech Republic}

\date{\today}

\begin{abstract}
We report on a theoretical study of the commensurability oscillations
in a quasi-two-dimensional electron gas modulated by a unidirectional
periodic potential and subjected to tilted magnetic fields with a
strong in-plane component. As a result of coupling of the in-plane
field component and the confining potential in the finite-width
quantum well, the originally circular cyclotron orbits become
anisotropic and tilted out of the sample plane.  A quasi-classical
approach to the theory, that relates the magneto-resistance
oscillations to the guiding-center drift, is extended to this case.
\end{abstract}


\maketitle
\section{Introduction}
An isotropic quasi-two-dimensional electron system with circular Fermi
contours undergoes the Lifshitz phase transition \cite{Lifshitz} under
the influence of a strong in-plane magnetic field.  The topology of
Fermi contours changes. The deviation from the circular shape depends
on the magnetic field strength and the form of the confining
potential.  In systems with two occupied subbands the excited subband
is emptied at a certain critical in-plane field, and the corresponding
second Fermi loop disappears.

The Fermi contours acquire an asymmetric egg-like shape in an
asymmetric triangular potential at the
hetero-interface.\cite{Heisz1993, Smrcka.JP.1993} In wide quantum
wells and double wells with a single occupied subband the Fermi
contours resemble the Cassini oval: as the in-plane magnetic field
increases the elongated convex curves acquire the concave peanut-like
shape.  At high in-plane fields the single Fermi line is split into
two parts.\cite{Lyo.PRB.95,Smrcka.JP.1995}

The deformation of a Fermi contour shape can be characterized by a
single experimentally measurable quantity, the
magnetic-field-dependent cyclotron mass.\cite{Smrcka.JP.1993,
  Smrcka.JP.1995} Its field-dependence can be studied e.g.\ by the
cyclotron resonance in the infrared region of the optical spectra
\cite{Kuriyama.SSC.1999, Takaoka.PhysB.2001, Takaoka.PhysE.2001,
  Aikawa.PhysE.2002, Marlow.PhysB.98} or the temperature damping of
Shubnikov-de Haas oscillations. \cite{Smrcka.PRB.95,
  Schneider.PhysB.01, Hatke.PRB.12} The closely related magnetic-field
dependence of the density of states is reflected in a resistance
oscillation measured as a function of the in-plane magnetic field.
\cite{Jungwirth.PRB.1997,Makar.PRB.00}

More details about the size and shape of cyclotron orbits can be
gained from the magneto-electron focusing experiments and 
 the commensurability oscillations measurement.
 \cite{Ohtsuka.PB.1998,Oto.PE.2001}

The commensurability oscillations \cite{Weiss.EP.1989,
  Winkler.PRL.1989, Gerhardts.PRL.1989, Beenakker.PRL.1989} --
oscillations of the magnetoresistance, measured at low temperatures
and in a low perpendicular magnetic field in the presence of a weak
modulation potential -- are periodic in the inverse field. Their
period reflects the commensurability of the cyclotron orbit diameter
and the modulation period.

The first usage of the commensurability oscillations measurement in
strong in-plane magnetic fields was to confirm the distortion of the
Fermi contour to the egg-like shape.\cite{Oto.PE.2001} Two
experimental arrangements were examined --- with a lattice vector of
the unidirectional lateral superlatices either parallel or
perpendicular to the in-plane magnetic-field component.  The observed
results were consistent with the theoretical prediction.

The influence of the Fermi loop egg-like deformation on the chaotic
electron dynamics in a two-dimensional antidot lattice was studied both
theoretically and experimentally. Reasonable agreement between the
theory and the experiment was
achieved.\cite{Soto.Micro.00,Soto.PRB.02}

Recently, the commensurability oscillations were investigated in
detail in a wide double hetero-junction well with an occupied bonding
subband and a unidirectional modulation
potential.\cite{Kamburov.PRB.2013} The caliper dimensions of the
in-plane field distorted Fermi contours obtained from the experimental
data were compared with the results of the first-principle
self-consistent calculation.  An overall semi-quantitative agreement
was achieved between the experimental and the theoretical results.
However, a systematic discrepancy was found between the observed and
the calculated elongation of the Fermi contour for the case of a
lattice vector parallel to the in-plane magnetic field.

To shed light on this apparent discrepancy between the theoretical and
the experimental findings, we try to extend the quasi-classical
approach, that relates the magnetoresistance oscillations to the
guiding-center drift, to the case of cyclotron orbits which are
anisotropic and tilted out of the sample plane.
\section{Cyclotron orbits in tilted magnetic field}
Let us consider a single bonding subband of a symmetric wide well or a
double well.\cite{Smrcka.JP.1993, Smrcka.JP.1995} Assuming the
in-plane magnetic field parallel to $y$-axis, the Fermi contour in the
$k_x-k_y$ plane is described by an expression
\begin{equation}
E_F = E(k_x) +\frac{\hbar^2 k_y^2}{2m},
\end{equation}
where $E_F$ denotes the Fermi energy. Only the $k_x$-dependence
of the energy is influenced by the in-plane field, $B_y$, while the
harmonic dependence of the energy on the wave vector component $k_y$
remains untouched. With increasing $B_y$ the curvature of $E(k_x)$
decreases for $k_x$ close to $k_x=0$ and, at a certain value of $B_y$,
becomes negative. A local maximum develops at $k_x=0$, accompanied
by two new minima positioned symmetrically around it.  The
corresponding Fermi contours resemble the Cassini ovals: the convex
curves elongated in $k_x$-direction acquire the concave peanut-like
shape at higher fields with the width of a peanut `waist' at $k_x=0$
shrinking to zero as $B_y$ increases.

The quasi-classical theory predicts that in a weak perpendicular
magnetic field $B_z$ an electron is driven around a Fermi contour by
the Lorentz force with the velocity components given by $v_x=1/\hbar\,
\partial E(k_x)/\partial k_x$ and $v_y=\hbar k_y/m$.  The related
real-space cyclotron orbits are similar in shape, but scaled by
$\ell_z^2 = \hbar/(|e|B_z)$ and rotated by $\pi/2$, as $v_x =
\ell_z^2\, d k_y/\! dt$ and $v_y= -\ell_z^2\, dk_x/\! dt$.

From these expressions we can calculate the in-plane field-dependent
period of the cyclotron motion $T$, or equivalently the cyclotron
frequency $\omega_c=|e|B_z/m_c$, where the cyclotron effective mass
$m_c$ is related to the shape of the Fermi contour by
\begin{equation}
\label{mc}
m_c = \frac{\hbar^2}{2\pi}\oint \frac{d k}{|\nabla_kE|} =
      \frac{\hbar}{2\pi}\oint \frac{d k}{\sqrt{v_x^2+v_y^2}}. 
\end{equation}
From here  we obtain the density of states $g$:
\begin{equation}
\label{g}
g= \frac{m_c}{\pi\hbar^2}.
\end{equation}
Eq.(\ref{mc}) also determines the relation between the electron
concentration  $N$ and the Fermi energy $E_F$ through
$N=g E_F$. In zero in-plane field this relation reduces to
$E_F=\hbar^2 k_F^2/2 m$, where the Fermi wave vector $k_F$ is given by 
$k_F=\sqrt{2\pi N}$.

Thus, assuming that the coordinates of a guiding center in $x-y$ plane
are $X$ and $Y$, the time dependence of an electron motion along
the real-space orbits is described by equations
\begin{equation}
\label{coor}
x(t)  = X + \ell_z^2 k_y(t),\,\,\,
y(t) = Y - \ell_z^2 k_x(t).
\end{equation}
Due to the finite well width  $d$  the real-space orbit has also a
$z$-coordinate:\cite{Smrcka.JP.1995}
\begin{equation}
\label{coorz}
z(t)  = \ell_y^2 \left(k_x - \frac{\hbar v_x}{m}\right),
\end{equation}  
where  $\ell_y^2 = \hbar/(|e|B_y)$.

Two new important features appear when compared with previously
discussed zero-in-plane-field cyclotron orbits. 

First, the velocity of an electron moving along an orbit is no longer
constant and depends on its position  on the Fermi
contour. Most pronounced changes go for the so called turning points,
i.e., the points on the Fermi contour where one of the velocity
components drops to zero. While in the zero-field case there were only two
pairs of turning points at crossings of the Fermi contour with the
$k_x$ and $k_y$ axes, new turning points appear when the Fermi contour
becomes concave. Their position on a Fermi line is related to
field-induced new minima of $E(k_x)$.

Second, as follows from Eqs.(\ref{coor},\ref{coorz}) the cyclotron
orbit is tilted. The $z$-coordinate of an electron does not
remain in the center of a well, and the electron moves from one side of a
well to the other when traveling along a real space path.

\section{Commensurability oscillations}
The quasi-classical description of the commensurability oscillations
-- the magneto-resistance oscillations due to the presence of a weak
unidirectional potential -- is based on the solution of the Boltzmann
equation in the relaxation time
approximation\cite{Beenakker.PRL.1989}.  The oscillations are periodic
in $1/B_z$ and their amplitude is proportional to the mean square
drift velocity averaged over the positions of the guiding centers. The
drift velocity of a given guiding center can be calculated by the
time-average of an oscillating electric field seen by an electron
traveling around the center along the real-space cyclotron orbit.

In zero in-plane magnetic field the cyclotron orbits are circles with
a radius $R=\ell_z ^2\, k_F$. If we assume a sinusoidal potential with
the lattice vector ${\bf a}$ oriented along the $y$-axis, the decisive
contributions to the drift velocity are obtained at the two turning
points $Y+R$ and $Y-R$, where an electron stops for a while, before
the velocity component $v_y$ changes sign.

It was shown\cite{Weiss.EP.1989} that the frequency $f$ of
commensurability oscillations is proportional to the radius of the
Fermi contour $k_F$ by
\begin{equation}
f  = \frac{2\hbar}{|e|a}k_F,
\label{freq}
\end{equation}
 i.e., the frequency $f$ measures
the caliper dimension of the Fermi contour.

In the following we discuss why the above formula cannot  be simply
extended to the case of non-circular Fermi contours by replacing
$k_F$  by the Fermi contour calipers, which  determine the position of the 
turning points on the real space orbits.

In finite $B_y$, the   time-averaged electric field of the orbit with
the guiding center coordinate $Y$ is given by 
\begin{equation}
\mathcal E(Y)=
\int_0^T \mathcal E_0 \bigl(z(t)\bigr) 
cos\left(\frac{2\pi}{a} y(t) \right) dt.
\label{drift1}
\end{equation}
The corresponding drift velocity reads
$v_\textrm{drift}=\mathcal E(Y)/B_z$,
the dependence of $y$ and $z$ coordinates on time is described through the
time dependent $k_x$ by Eqs.(\ref{coor},\ref{coorz}). 

When compared with the zero-field case, the main difference stems from
the following facts.

\begin{center}
\begin{figure}[htb]
\includegraphics[width=\linewidth]{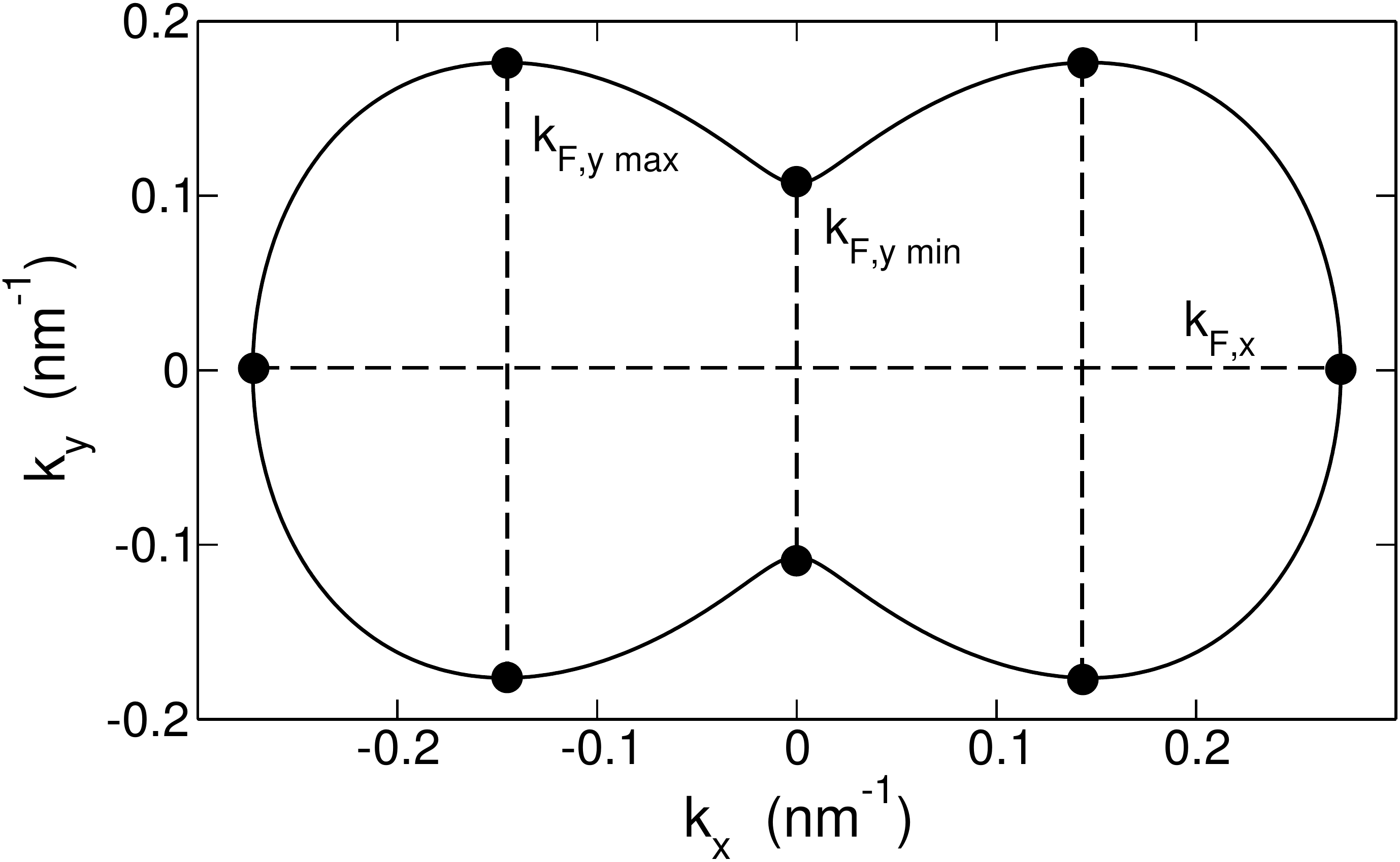}
\caption{The caliper dimensions of the concave Fermi contour at $B_y
  =8\, \textrm{T}$.  Dots mark the turning points on the related
  cyclotron orbits  rotated by $\pi/2$ and multiplied by $\ell_z^2$.}
\label{kxky8}
\end{figure}
\end{center}

There are four more turning points for concave Fermi
contours where the velocity
components drop to zero, as  shown in the Fig.(\ref{kxky8}).

Moreover, we assume that the source of the unidirectional
periodic modulation is only on one side of a quantum well and 
therefore the modulation potential is partly screened by the
quasi-two-dimensional electron layer of a finite width.  Then the
strength of the oscillating electric field is $z$-dependent and need
not be the same on both sides of a well.

As mentioned above, the transport coefficients obtained by the
solution of the Boltzmann equation, or from the linear response
theory,\cite{Gerhardts.PRL.1989,Winkler.PRL.1989} depend on the mean
square of the drift velocities. It was shown by both methods that
${\Delta \varrho_{yy}}$, the change of the magneto-resistance in the
presence of a weak unidirectional potential, is proportional to the
mean square drift velocity averaged over $Y$,
\begin{equation}
\frac{\Delta \varrho_{yy}}{\varrho_0}\, \propto \,
\frac{\omega_z^2\tau^2}{\sigma_0} \frac{1}{a} \int_0^a v_\textrm{drift}^2 (Y) dY
=\frac{\omega_z^2\tau^2}{\sigma_0} \langle v_\textrm{drift}^2 \rangle.
\label{drift2}
\end{equation}
Here $\tau$ is the relaxation time, $\varrho_0$ and $\sigma_0$ are
the  resistivity and conductivity at $B_z=0$, respectively.

Similar expressions can be written for the lattice vector oriented
along the $x$-direction.

\section{Model calculation}
To illustrate the relevance of the above extension of the
quasi-classical description of commensurability oscillations we
present the results of a numerical calculation based on the simple
tight-binding model of a double well.\cite{Hu.PRB.92, Lyo.PRB.94,
Kurobe.PRB.94} This model captures quite well  the essence of physics
of double wells and wide quantum wells subjected to a strong in-plane
magnetic field.

We consider two strictly two-dimensional electron layers in very
narrow quantum wells separated by a barrier. The model is
characterized by two parameters, the interlayer distance $d$ and the
hopping integral $t_h$.  Such a system represents the first step from
strictly two-dimensional to three-dimensional structures.

\begin{center}
\begin{figure}[h]
\includegraphics[width=\linewidth]{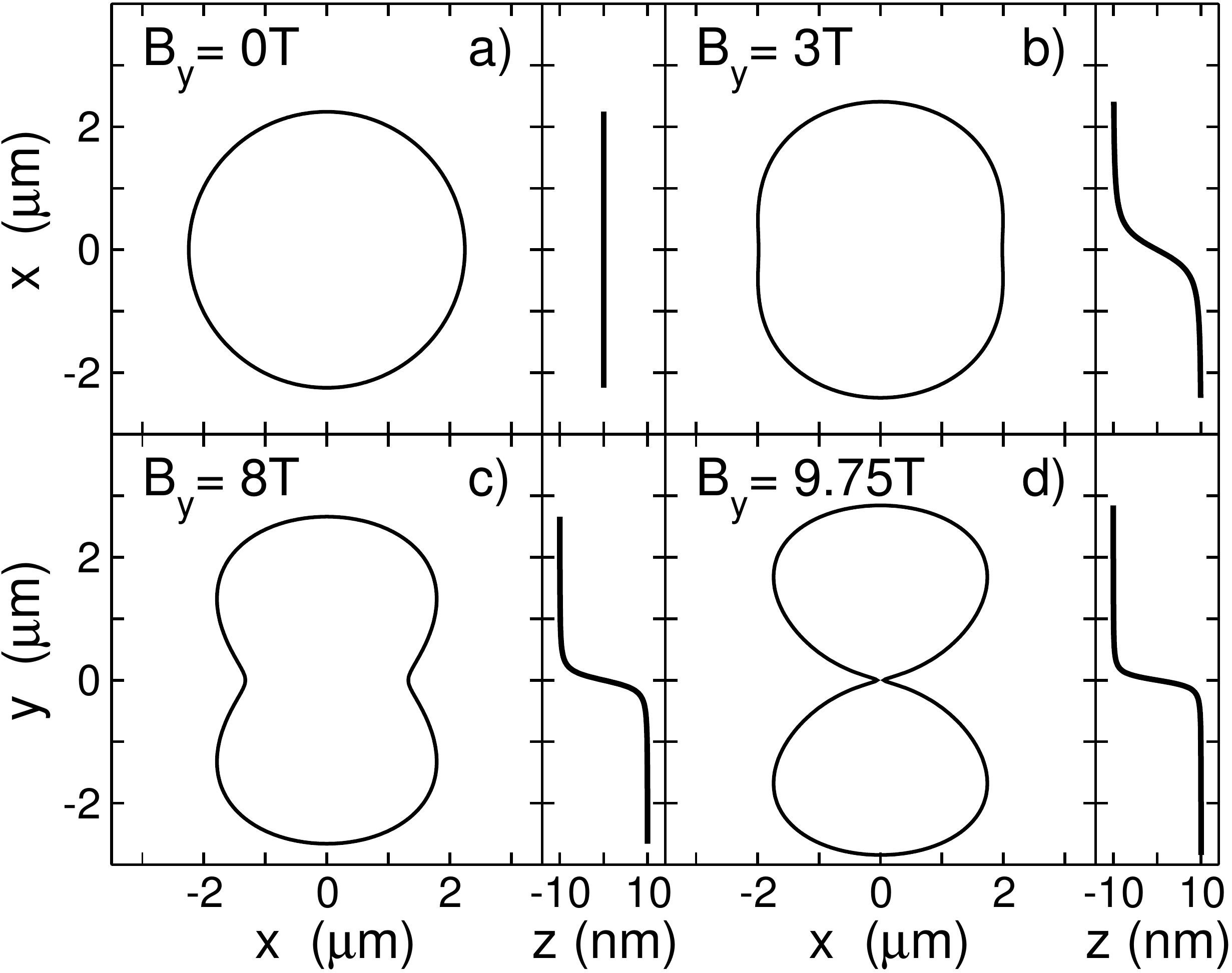}
\caption{Projections of cyclotron orbits on the $x-y$ and $y-z$
  planes, $B_z \approx 0.07\, \textrm{T}$.  a) $B_y =0\,
  \textrm{T}$. The orbit is circular and lies in the middle of a
  double well at $z=0$.  b) $B_y =3\, \textrm{T}$. The orbit is on
  the border between the convex and concave curve. The $z$-coordinate
  changes sign at $y=0$. c) $B_y =8\, \textrm{T}$. The concave shape
  of the orbit is fully developed.  An electron spends most time near
  the opposite sides of the structure at $z=\pm d/2$. d) $B_y =9.75\,
  \textrm{T}$. The orbit is close to splitting. The change of sign of
  the $z$-coordinate near $y=0$ is very sharp.}
\label{FContours}
\end{figure}
\end{center}

The diagonalization of a $2\!\times\!2$ matrix yields the
$k_x$-dependence of the bonding subband energy in the form
\begin{equation}
E(k_x) =\frac{\hbar^2 k_x^2}{2m}+\frac{e^2 d^2 B_y^2}{8
  m}-\sqrt{\delta^2+t_h^2},
\end{equation}
where $\delta$ is the abbreviation for $\hbar |e| d B_y k_x/2m$.  The
Fermi energy $E_F$ corresponding to the fixed concentration of
carriers $N$ is determined by Eqs. (\ref{mc},\ref{g}). 
Here we choose $N=8 \times 10^{11} \, \textrm{cm}^{-2}$, $d=20\,
\textrm{nm}$ and $t_h=2\, \textrm{meV}$. 
A few examples of cyclotron orbits calculated based on these
parameter are presented in the Fig.\ref{FContours}.

The $B_z$-dependence of commensurability oscillations is given by
Eqs.(\ref{drift1}) and (\ref{drift2}). These expressions were
evaluated by numerical integration for the fixed in-plane field
components $B_y$ between $0$ and $9.75\, \textrm{T}$. The domain of
$B_z$ was chosen between $0.05$ and $0.35\,\textrm{T}$. This is
consistent with a range of fields commonly used in experiments --
below the minimum field the oscillations are usually damped, above the
maximum field the Shubnikov-de Hass oscillations dominate. Note that
this choice is not an inherent property of the quasi-classical
approach.  But, as it will be shown later, it can slightly influence 
interpretation of our model calculation.

To test the accuracy of our numerical method, besides of $\langle
v_\textrm{drift}^2 \rangle$ we also calculate the averaged drift
velocity $\langle v_\textrm{drift} \rangle$. It defines the
macroscopic current $e\langle v_\textrm{drift} \rangle$ through the
sample which must, of course, disappear in the thermodynamic
equilibrium.

\begin{figure}[h]
\includegraphics[width=\linewidth]{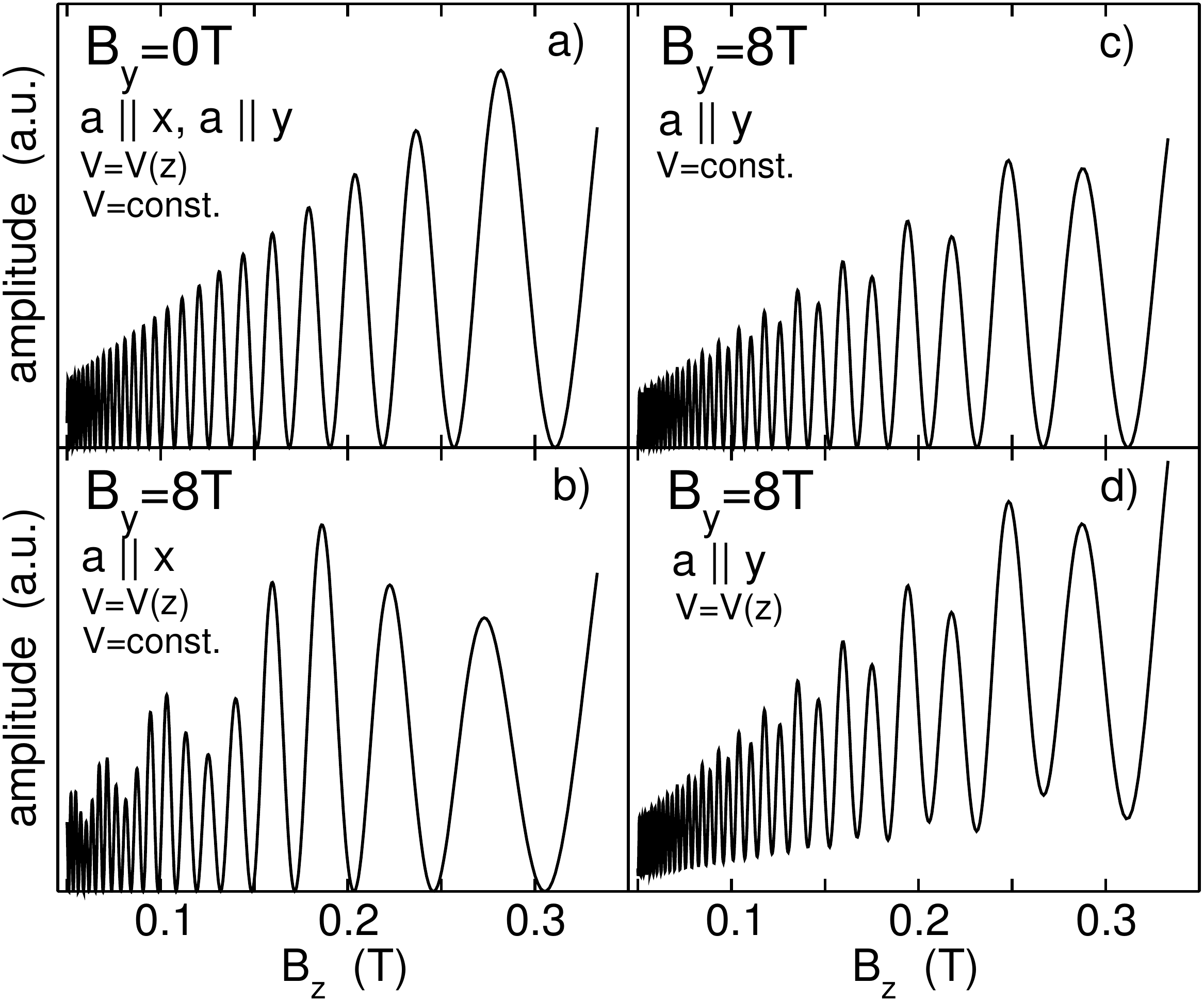}
\caption{Commensurability oscillations as a function of $B_z$.  a)
  $B_y =0\, \textrm{T}$. The orbit is circular and the same form of
  oscillations for ${\bf a} \!\parallel\!x$ and ${\bf a\!}\parallel\!
  y$ is obtained. The $z$-dependence of a modulation potential plays
  no role.  b) $B_y =8\, \textrm{T}$, ${\bf a}\!\parallel\!x$. Again,
  the $z$-dependence of a modulation potential plays no role, as an
  electron feels the same electric field at $x$ and $-x$. c) $B_y =8\,
  \textrm{T}$, ${\bf a}\!\parallel\!y$ and a $z$-independent
  modulation potential. Two frequencies of oscillations are resolved,
  their aplitude reaches zero.  d) $B_y =8\, \textrm{T}$, ${\bf
    a}\!\parallel\!y$ and a $z$-dependent modulation potential. An
  electron feels different electric fields at $y$ and $-y$. The
  amplitude of oscillations does not drop to zero. The weights of two
  frequencies in the frequency spectrum are changed.}
\label{oscil}
\end{figure}

We first test our model on two examples: the standard
circular Fermi contour at $B_y=0\, \textrm{T}$ and a strongly
anisotropic Fermi contour at $B_y=8\, \textrm{T}$. The output of our 
model calculation is presented in the Fig.\ref{oscil}.

The results for a circular Fermi contour are shown in the
Fig.\ref{oscil}\! a).  They agree with results published previously, 
the oscillations reach zero in each period and
their frequency satisfies the commensurability condition $f=
\frac{2\hbar}{|e|a}\,k_F$, given by Eq.(\ref{freq}). The period of
oscillations is the same for a lattice vector ${\bf a}$ oriented along
the $x$ and $y$-axes. The reason is that an electron traveling along
the circular orbit stays all the time in the middle of the structure,
$z = 0$, and, therefore, the $z$-dependence of a modulation potential
plays no role.

The case of  anisotropic Fermi contours is quite different.  The
dependence of oscillations on the lattice vector orientation is strong
and the $z$-dependence of the modulation potential can play an important
role. 

Let us first consider the lattice vector ${\bf a}$ parallel to the
$x$-axis. As in the previous case, $B_y =0\, \textrm{T}$,
the $z$-dependence of the modulation potential has almost no influence
on the resulting shape of commensurability oscillations, shown in the
Fig.\ref{oscil}\! b). The reason is that the $z$-coordinate depends only
on $k_x$ and an electron feels the same electric field at $x$ and
$-x$.  The fast Fourier transform (FFT) of oscillations yields three
frequencies related to four turning points which correspond to
$\pm\, k_{F,y\textrm{ max}}$, two turning points corresponding to
$\pm\, k_{F,y\textrm{ min}}$, and their average.

Now we turn our attention to ${\bf a}\!\parallel\! y$ and a
$z$-independent modulation potential. The calculated results are
presented in the Fig.\ref{oscil} c). The FFT of oscillations yields
two frequencies close to $f_1 =\frac{2\hbar}{|e|a}\, k_{F,x}$ and $f_2
= \frac{\hbar}{|e|a}\,k_{F,x}$. As mentioned above their deviations
from the exact values and their weights in the frequency spectrum
depend slightly on the chosen domain of $B_z$.  The frequency $f_1$ is
related to the turning points $\pm\,k_{F,x}$ shown in Fig.\ref{kxky8}.
The frequency $f_2$ appears for the first time at the critical field
$B_y$ for which the Fermi contour becomes concave, in our case at
$B_y=3\,\textrm{T}$. The weight of $f_2$ in the frequency spectrum
increases with increasing $B_y$ and reaches maximum at a critical
field $B_y= 9.75\,\textrm{T}$ above which the Fermi contour splits.
The reason is that at this field both velocity components $v_x$ and
$v_y$ drop to zero at ${\bf k} =0$. Note that the period of cyclotron
motion $T$ and the cyclotron mass $m_c$, given by Eq.(\ref{mc}), also
have a singularity for the same reason here.

Next we consider ${\bf a}\!\parallel\! y$ and a $z$-dependent
modulation potential.  The $z$-dependence of the modulation affects
mainly the role of a real-space trajectory in the vicinity of the
turning points related to $k_{F,x}$ and $-k_{F,x}$. According to
Eq.(\ref{coorz}) the $z$-coordinate depends on $k_x$ and an electron
sees different electric fields near these two points.  As a
consequence, the electric field averaged over the cyclotron orbit is
modified. The amplitude of oscillations does not drop to zero, as can
be seen in the Fig.\ref{oscil} d), and the weights of frequencies
$f_1$ and $f_2$ in the frequency spectrum change.

\begin{figure}[h]
\includegraphics[width=\linewidth]{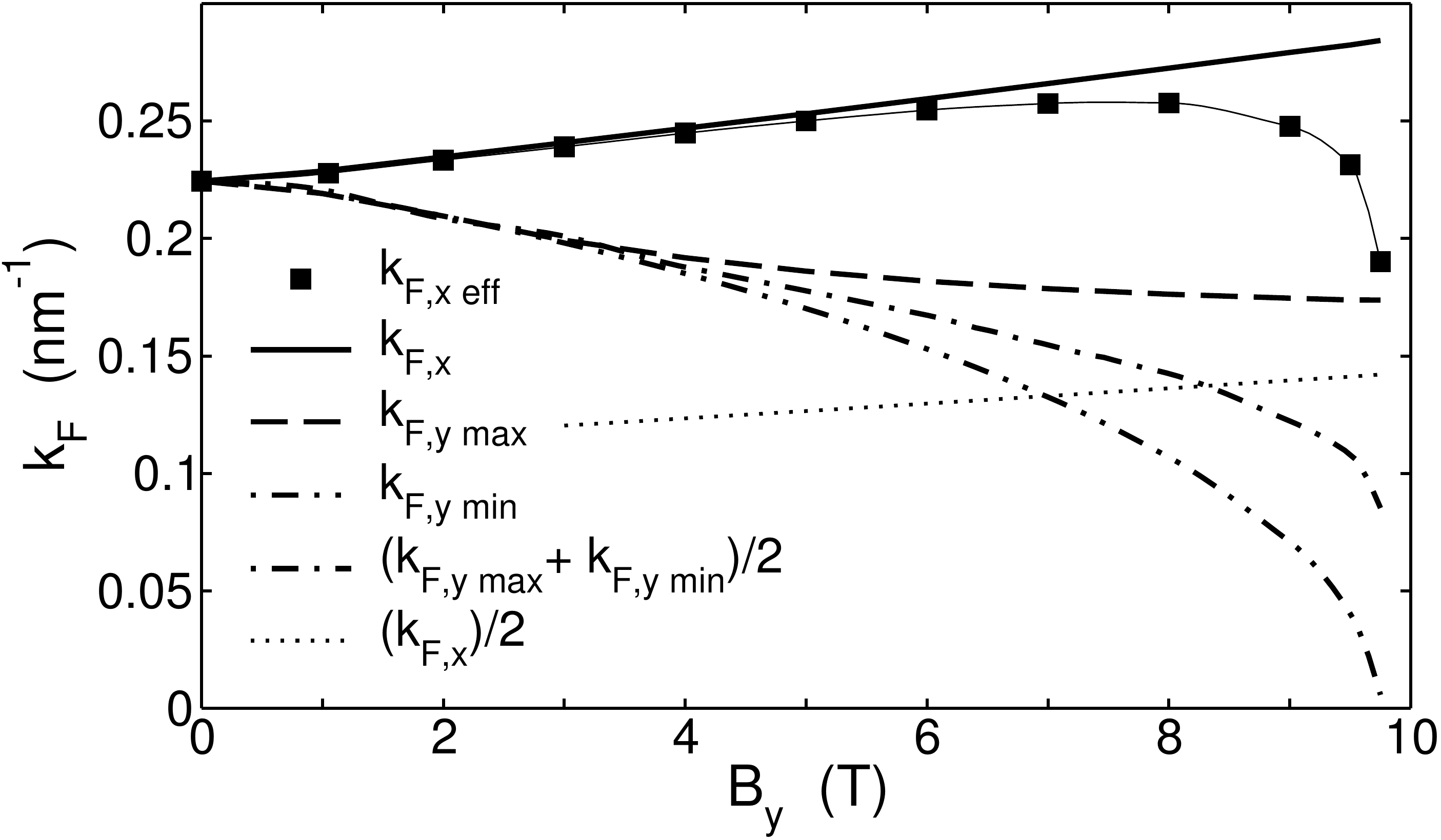}
\caption{The in-plane field dependence of the Fermi contour caliper
  dimensions. For ${\bf a}\parallel x$ the $k_{F,y\textrm{ max}}$
  splits into $k_{F,y\textrm{ min}}$ and $(k_{F,y\textrm{ max}} +
  k_{F,y\textrm{ min}})/2$ above the critical field $B_y$ for which
  the Fermi contour becomes concave.  For ${\bf a}\parallel y$, the
  value $k_{F,x\textrm{ max}}/2$ appears abruptly at this critical
  field. An effective $k_{F,x\textrm{ eff}}$ is derived from the
  weighted average of frequencies $f_1$ and $f_2$ corresponding to
  those vectors.}
\label{Fig4}
\end{figure}

The main results of our model calculations are shown in the
Fig.\ref{Fig4}.  The evolution of the Fermi contour calipers in the
in-plane magnetic field $B_y$ is presented together with the values
derived from the FFT of commensurability oscillations.  The
oscillations were calculated assuming a $z$-dependent modulation
potential that yielded the reduced electric field in the second
layer. Here $\mathcal E(-d/2)=0.4\times \mathcal E(d/2)$ was chosen.

Our calculation does not take into account the line broadening by the
inaccuracy of instruments, by the finite temperature and/or by
electron scattering. On the other hand, the line broadening strongly
influences interpretation of experimental data.  Even in the most
recent measurements \cite{Kamburov.PRB.2013} only one frequency was
resolved in oscillation spectra measured both for ${\bf a}\parallel x$
and ${\bf a}\parallel y$. For the case of ${\bf a}\parallel x$ the
good agreement with the frequency corresponding to $ k_{F,y\textrm{
    max}}$ was found, for ${\bf a}\parallel y $ the measured frequency
is between $f_1$ and $f_2$ mentioned above.

In our calculation, which does not take into account the line
broadening, the agreement between the Fermi contour calipers and the
values derived from the FFT is very good for ${\bf a}\parallel
x$. Therefore only the curves corresponding to contour calipers are
plotted in the Fig.\ref{Fig4} as functions of $B_y$. The
$k_{F,y\textrm{ max}}$ splits into $k_{F,y\textrm{ min}}$ and
$(k_{F,y\textrm{ max}} + k_{F,y\textrm{ min}})/2$ above the critical
field $B_y$ for which the Fermi contour becomes concave.  The weight
of the frequency related to $k_{F,y\textrm{ max}}$ dominates.

The case of ${\bf a}\parallel y$ is different. For $B_y$ below the
critical value only the peak related to $f_1$ is observed in the FFT
spectrum. The second peak corresponding to $f_2 =f_1/2$ appears with a
very small weight at the critical field and dominates at high fields.
To mimic the experimental results, which resolve only one frequency,
we plot an effective $k_{F,x\textrm{ eff}}$, derived from the weighted
average of frequencies $f_1$ and $f_2$, except of the values based on
the frequencies $f_1$ and $f_2$ themselves.

\section{Summary and conclusions}
We have studied the commensurability oscillations in a
quasi-two-dimensional electron gas modulated by a unidirectional
periodic potential and subject to tilted magnetic fields with a strong
in-plane component.  A quasi-classical approach to the theory, that
relates the magneto-resistance oscillations to the guiding-center
drift was employed.  A systematic discrepancy between the observed and
the calculated elongation of the Fermi contour for the in-plane
magnetic field parallel to the lattice vector, found in\,
\cite{Kamburov.PRB.2013}, is explained by the in-plane magnetic-field
distortion of the cyclotron orbits .

Our arguments are as follows: First, the velocity of an electron
moving along an orbit is no longer constant and depends on its
position on the Fermi contour.  Second, the cyclotron orbit is tilted,
the electron moves from one side of a well to the other, and sees the
different strengths of the one-side-modulation potential on the
different well sides.

The numerical simulation is based on the simple tight-binding model of
a double well, the spin of electrons is not taken into account.

\section{Acknowledgements}
This work was supported by grant Nr. LM2011026 of the Ministry of
Education of the Czech Republic.  J.\ Ku\v{c}era is ackowledged for his
careful reading of the manuscript.

\end{document}